**Material Challenges for Colloidal Quantum Nanostructures in Next Generation Displays**

Yossef E. Panfil[1,2], Meirav Oded[1,2], Nir Waiskopf[1,2], Uri Banin[1,2]*

1 Institute of Chemistry, The Hebrew University of Jerusalem, Jerusalem 91904, Israel.

2 The Center for Nanoscience and Nanotechnology, The Hebrew University of Jerusalem, Jerusalem 91904, Israel.

* Corresponding author. Email: uri.banin@mail.huji.ac.il (U.B.)

**Abstract**

The recent technological advancements have greatly improved the quality and resolution of displays. Yet, issues like full color gamut representation and long lasting durability of the color emitters require further progression. Colloidal quantum dots manifest an inherent narrow spectral emission with optical stability, combined with various chemical processability options which will allow for their integration in display applications. Apart from their numerous advantages, they also present unique opportunities for the next technological leaps in the field.

**1. Introduction**

Displays are all around us. They are in our smartphones, laptops, TVs, and cars and they have become intwined in our everyday life with new features being added, such as the Augmented Reality technology. The onset of our present age, the "Information Age", is to a great extent associated with the technological advancements of displays, which are the medium that enables this era. This is even more expressed with the development of the Covid-19 pandemic. What would we do without the virtual meetings and classes, afforded only by the development of recent technology?

The evolution and revolution of display technologies were always closely associated with advancements in materials chemistry. The cathode ray tube (CRT) display was the workhorse of display technology for many years before the arrival of plasma screens and later on liquid crystal display (LCD) and organic light-emitting diode (OLED) displays. Nonetheless, the search for a lifelike vision experience is still ongoing, and striving for constant improvement in performance such as low energy consumption, better colors, thinner formats and also new functionalities such as transparent and flexible displays, all of which hinge upon state-of-the-art materials developments.

Modern displays contain millions of pixels which are subdivided into red, green, and blue subpixels. The fine control of each subpixel determines the color which the human eye perceives from this pixel. Figure 1a depicts the International Commission on Illumination (CIE) chromaticity diagram. This diagram span the colors which are visible to the human eye in terms of hue and saturation. Any three basis colors located on the map defines a triangle in which the enclosed colors can be generated by a combination of its corners. The key to enlarging the area of the above triangle is to purify the basis colors. Practically, this means that as the red, green, and blue subpixels will demonstrate a narrower spectrum with no "cross-talk" between the subpixels, the display will envisage a wider range of colors for the human eye.

Colloidal quantum nanostructures, of which the Colloidal Semiconductor Nanocrystal (SCNC) Quantum dots is an archetypical example, are promising materials for displays and have already proven their relevance for greatly improving display technologies. The SCNCs are tiny crystals covered by passivating ligands on their surface, which are smaller than the exciton Bohr radius of the semiconductor material. Upon absorption of a photon with energy higher or equal to the band edge or upon charge injection, an electron-hole pair (exciton) is generated. This exciton is lasting for tens of nanoseconds before recombining to give a photon with a color corresponding to the band edge energy. These nanocrystals emerged at the beginning of the 1980s and since then had been developed and studied extensively for four decades now[1]. Due to the combination of their unique size, shape, and composition dependent properties, alongside their facile bottom-up fabrication by wet chemical means and ability to manipulate them from solution flexibly, SCNCs manifest inherent advantages as emitters in displays.

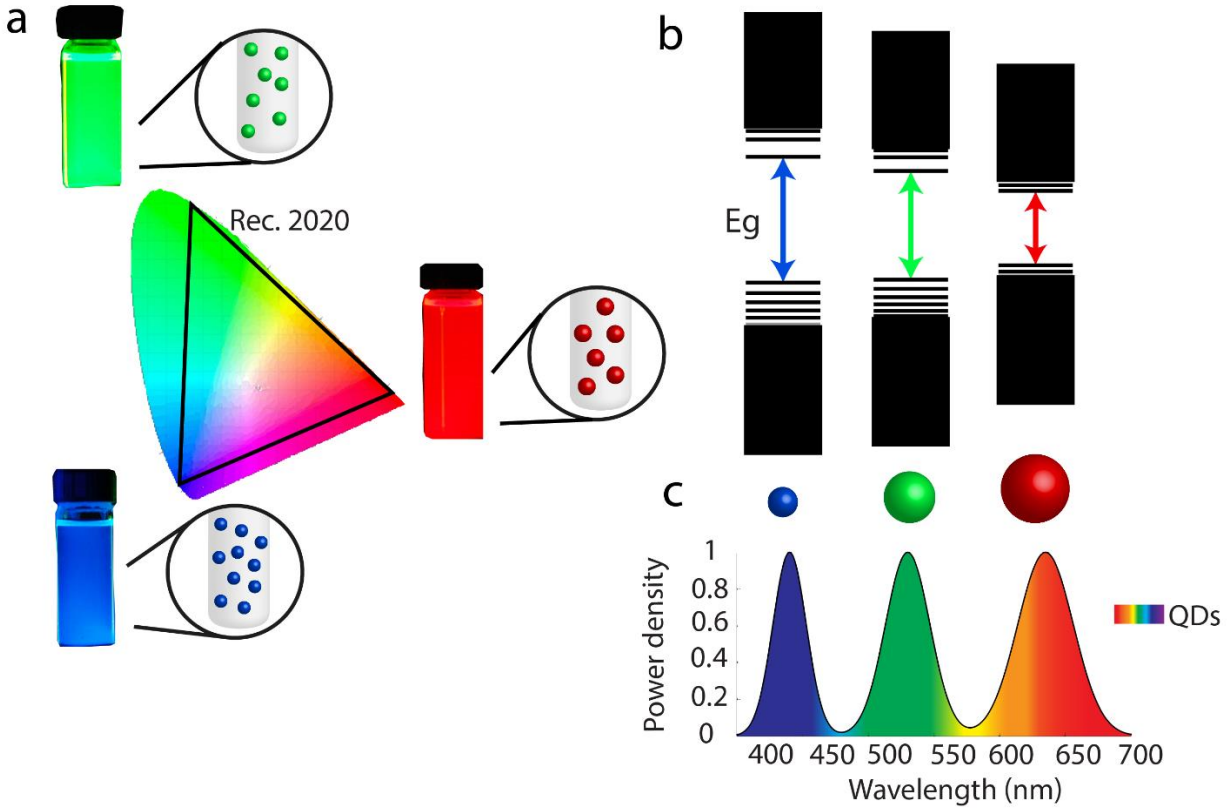

**Figure 1.** *The advantages of SCNCs for display applications.* a) 1931 CIE chromaticity diagram with the Rec. 2020 color gamut (black triangle) together with red, green, and blue vials with illustrations of large (red emitting), medium (green emitting), and small (blue emitting) particles, respectively. b) Schematic representation of the electronic energy structure of different-sized SCNCs, dictated by the quantum confinement effect. c) Typical emission spectra of red, green, and blue SCNCs.

The emission wavelength of QDs is tunable and depends simply on the size of the nanocrystals. As an example, CdSe nanocrystal at the size of ~2nm emits blue light, whereas CdSe at the size of ~6nm will emit red light (Fig. 1a-b). Therefore, colloidal quantum dots prepared by a controlled wet chemistry synthesis can span all the visible range and in the context of displays, red, green, and blue emitting nanocrystals can be synthesized. Moreover, the fine tuning of the three basis wavelengths by the nanocrystal size can broaden the color gamut coverage as opposed to e.g. phosphors materials in which the emission wavelength is given by the characteristics of the specific material.

Along with the emission wavelength tunability, an additional important characteristic which makes SCNCs so appealing to display technology is their narrow emission spectrum. A batch of SCNCs will have a smaller than 30nm full width at half maximum (FWHM) peak distribution (Fig. 1c). This characteristic is attributed, firstly, to Quantum confinement effects which leads to discrete energy levels both in the conduction and valence band of the nanocrystal (Fig. 1b), and secondly, to the mature and well controlled wet chemistry synthesis allowing narrow nanocrystals size dispersion. Narrow emission spectrum in the red, green, and blue is essential for increasing the color coverage to achieve the Rec. 2020 color gamut standard.

The emitters inside a display must be also efficient, bright, and stable. A major leap towards bright and stable SCNC was the introduction of a CdSe/ZnS core/shell nanocrystal[2]. In this heterostructure, the core material is covered by a wide band-gap material which serves as a potential barrier between the core, where the charge carriers reside, to the outer surface which can contain trap states. This concept has led to the production of SCNCs with fluorescent Quantum Yield (QY) close to 100%[3].

Besides the above mentioned obvious advantages of SCNCs as emitters in displays, SCNCs with different dimensionalities such as quasi 1-dimensional nanorod, dot-in-rod, and quasi 2-dimensional nanoplatelets have been developed. These SCNCs hold special properties such as polarized emission or narrow spectrum which can improve the display technology as will be elaborated below.

As discussed - SCNCs are nowadays stable, bright, and efficient emitters which can revolutionize the way we perceive the world throughout screens. In this review, we will start with the introduction of different modalities in which SCNCs are being used today and in the near future in display technology. We will continue and cover the broad pallet of SCNCs and their inherent advantage of being chemically processable for pixels patterning. Then we will discuss some major challenges of this technology, in finding heavy-metal free materials that will meet all the needs of nanocrystals in displays. Finally, we will outlook at the use of SCNCs in multi-functional and new suggested modalities of displays.

**2. Road map for NCs in displays**

As mentioned in the introduction, an important characteristic of SCNCs for display technology is their narrow emission spectrum. Indeed, the first attempts to incorporate SCNCs in displays were by utilizing this advantage in liquid crystal displays (LCDs). The basic architecture of an LCD is a white backlight which is then attenuated or even switched off by applying a voltage on a liquid crystal cell, located between two polarizers, to vary its polarization, such that the light either passes the top polarizer, (pixel on) or blocked (pixel off). Then, the light passes through red, green, and blue color filters while the ratios between the intensities of the three subpixels can span any color composed out of the three base colors.

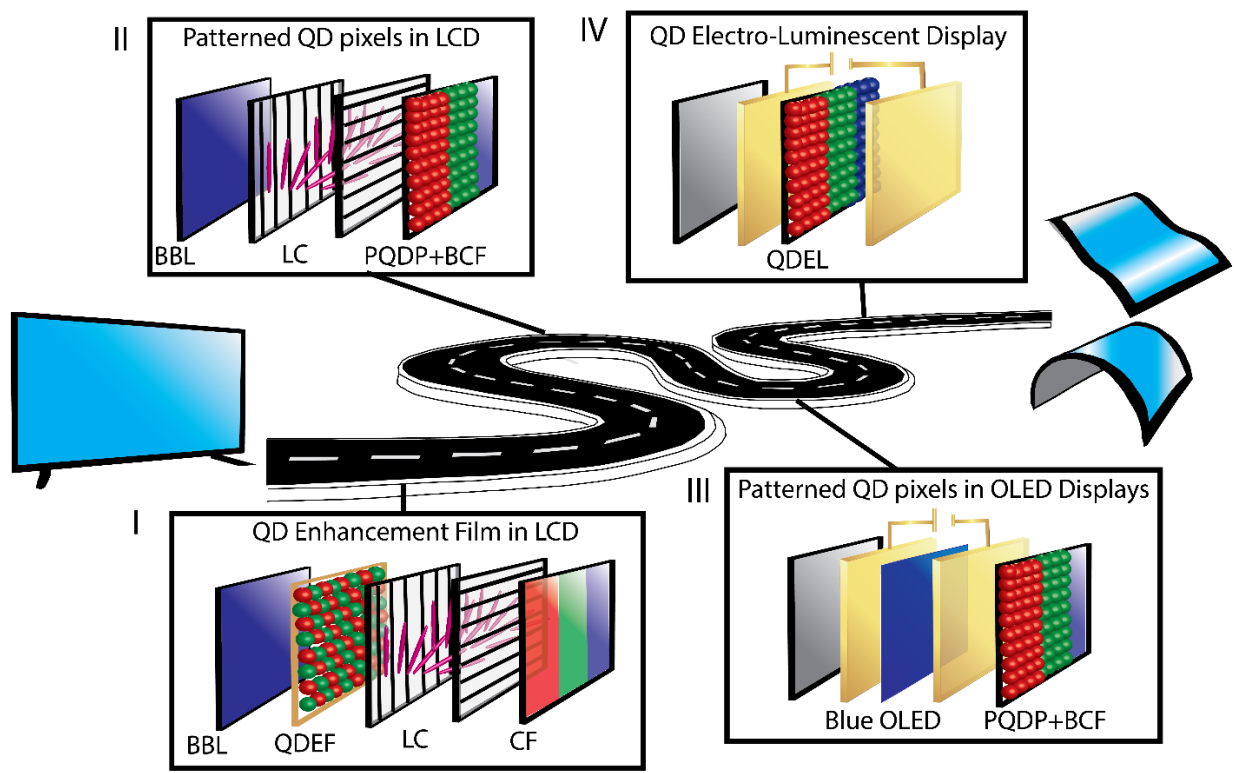

**Figure 2.** *The roadmap to SCNC QD displays.* I) QD enhancement film in LCD. In which BBL is blue backlight, QDEF is QD enhancement film, LC is liquid crystal, and CF is color filter. II) Patterned QD pixels in LCD. In which PQDP+BCF is patterned QD pixels+blue color filter. III) Patterned QD pixels in OLED displays. IV) QD electro-luminescent display. In which QDEL is QD electro-luminescence.

The first stage in the QD displays road map was to utilize a blue backlight along with red and green SCNCs (Fig. 2I). These QDs have strong absorption in the blue region of the spectrum and their red and green emission are very narrow. In this way, the red and green spectra are dictated by the QDs and not by the color filters. This increases the color gamut coverage and also improves the brightness and efficiency of the displays because the QDs spectrum is narrower than the color filters transmission spectrum.

For this color conversion and enrichment application, there are three approaches to incorporate the SCNCs in the backlight of displays. The first one is to place the SCNCs directly on the blue back light emitting diode (LED) chips. In this way, the least amount of QDs is needed. However, in this configuration the SCNCs are very close to the blue LED where the temperature is very high in addition to the high optical flux. As a result, the lifetime and stability of the SCNCs are sacrificed. A second approach is called "on-edge". In this method, the QDs are not directly on the LED chip but are further away inside a tube between the blue LED and the light guide. In this way, a small amount of QDs is needed and the temperature and flux are acceptable. However, this configuration brings about mechanical integration problems. The third approach is more widely used. The red and green QDs are suspended in a polymer film that is placed after the light guide and it is used as a simple "drop-in" solution. These films are often called "Quantum Dot Enhancement Film" (QDEF). In this method, The QDs are at room temperature and the flux is low. However, the QDs consumption is higher.

Since colloidal QDs are easily manipulated by wet chemistry means, new ways of pixel patterning like "inkjet" or photolithography, which are able to produce less than 2 micron subpixels, are emerging as will be discussed in more detail below. This paved the way for the next step in QD displays, using patterned subpixels of SCNCs possibly alleviating the use of the color filters in LCDs altogether (Fig. 2II). By this, the efficiency and brightness of the displays can be dramatically enhanced. While for SCNCs placed in the back light unit, the mixed red, green, and blue light is going through the three passive color filters, leading to high losses and poor efficiency and brightness, for patterned SCNCs which are acting as active color filters, the losses are minimized since the blue back light is converted to red and green light by the red and green SCNCs subpixels. This configuration also leads to improved color accuracy at large viewing angles since the light is generated in the front of the display. In addition, since only the blue light is modulated by the liquid crystal cell, the thickness, response time, and driving voltages can be reduced, shortened, and decreased, respectively.

Organic Light Emitting Diode (OLED) displays were thought for a long time to be the future of TV screens technology because of the inherent limitation of LCD. LCDs suffer from narrow viewing angles, they waste energy, they struggle to show true black colors which inevitably leads to a limited dynamic range, the liquid crystal which is twisted by the electric field is limiting the switching speed which might be a problem for fast-motion content in sports and movies and from bulky shape because of the large number of layers needed in the LCD. An OLED display works without a backlight. It consists of an anode, hole injection and transport layers, an emitting layer of organic material, an electron transport and injection layer, and a cathode. Because of its structure, the OLED display can generate true black levels by nulling the electric current. It has a wider viewing angle and it can be made very thin and transparent while also allowing for flexible displays. The desire to combine the aforementioned increased color gamut that SCNCs can offer and their advantages, also to the emerging OLED displays, has led to the next step in the QD display road map. In this configuration, the QDs are used again instead of the color filters but this time no backlight is used. The patterned QDs are placed directly on top of the blue OLED subpixels (Fig. 2III).

Despite all of the abovementioned configurations, the ultimate QD display which is at the future point of the road map is an electro-luminescent QD display. This configuration is similar to the OLED architecture with electron and hole injection and transport layers but the emissive material in the layers structure is the QDs themselves (Fig. 2IV). The advantages of the electro-luminescent QD display configuration are vast. The stability and lifetime of these displays will be prolonged because of the inorganic materials in the emissive layer. The thickness of the displays will be minimized paving the way to flexible and transparent displays which can drive augmented reality applications, and true black levels together with extended color gamut coverage are expected. Moreover, unlike the configurations so far in which the SCNCs emission is generated by blue light excitation with a potential for unabsorbed blue light leakage in the green and red subpixels, in electro-luminescent QD display the excitons in each one of the subpixels are generated electrically, alleviating cross-talk between different colors.

Nonetheless, electro-luminescent QD displays have their challenges. Unlike optical excitation, in electrical injection, the charge balance depends on the layers structure which is not symmetric for the electrons and holes. As a result, the QDs are usually charged and are subjected to a nonradiative Auger process which is faster than the radiative process. Moreover, the emissive layer is subjected to a high electric field which can separate the electron and hole wavefunctions leading to decreased QY. In this context, it is worth mentioning two major developments that can partially solve the aforesaid problems. Core/shell nanocrystals with thick shell[4,5] and graded shell[6–9] were developed. These nanocrystals are known to have reduced Auger non-radiative recombination process, and together with a graded shell the electron and hole wavefunction separation under the electric field will be moderated.

In the next sections, we will dive deeper into the SCNCs chemistry and properties, and first we will discuss the vast material and dimensionalities that SCNCs have to offer and their advantages for display applications.

## 3. The Colloidal Quantum Materials palette
### a. Diverse compositions and dimensionalities

The most prominent feature associated with SCNCs is the ability to tune their emission wavelength with size, as was explained above. The variety in the composition and/or dimensionality of SCNCs provides additional handles for tailoring specific electrical and optical properties that meet the requirements for display applications. For example, core/shell structures, in which the NCs are composed of two (or more) separate areas of SC materials along their radial direction, allows for shaping the potential energy profile of the charge carriers (electron and hole). A composition in which the band alignments of both the core and shell SCs are straddling, are referred to as Type I systems. In these systems, the core is electronically passivated and both charge carriers are localized within the core. The core/shell interface is passivated by the SC shell which typically results in high photoluminescence QY and enhanced stability,[10] both are imperative traits for display applications.

When designing display-compatible core/shell systems, the band alignment consideration is not the only criteria to take into account. One must also relate to the dissimilarity between the lattice parameters of the SC materials composing the system. For SCs with close lattice parameters, successful passivation is usually afforded and structures with high QY are attained. However, in cases of large lattice mismatch between the core and the shell materials, during shell growth structural defects can occur at the interface which serve as traps, causing a reduction in the NCs' QY. One way to overcome this is by using a mediating layer of an alloy, or a graded shell which gradually changes its composition from the core to the shell SC materials. An example of the influence of the graded shell on the optical properties of the NCs was reported on CdSe/Cd$_{1-x}$Zn$_x$S-seeded nanorods, in which a graded rod-like (quasi one-dimensional, q-1D) shell of varying composition, Cd$_{1-x}$Zn$_x$S, was grown around a CdSe QD (zero-dimensional, 0D), generating a mixed-dimensional heterostructure with a type I band alignment.[8] Figure a presents scanning transmission electron microscopy (STEM) images with an elemental mapping of Zn, Cd, and Se, evidencing the dot-in-rod structure as well as the graded shell growth. The growth of the radially graded shell composition resulted in bright green emission with optical stability, manifested in minimal blinking of the fluorescence of single particles. Moreover, due to the elongated shape of the shell, the emission was highly polarized as well. Indeed, nanorods, as well as other 1D systems, exhibit intrinsic linearly polarized emission along their long axis, in contrast to the non-polarized emission of the spherical 0D structures.[11] This trait is also expressed in mixed dimensionality structures, of 0D-1D such as the dot-in-rod structure discussed above. Figure b demonstrates the photoluminescence of a film containing aligned CdSe/CdS dot-in-rod NCs, viewed through a polarizer. As the polarizer is aligned with respect to the NCs long axis (upper image) a bright red emission is observed, whereas for a polarizer in an orthogonal alignment to the NCs long axis (lower image), almost no emission is witnessed. Figure c displays a zoom-in polarized microscope image of the interdigitated electrode set-up used to align the nanorods. The property of q-1D and 1D NCs structures to emit polarized light plays an important role towards enhanced polarized displays. In LCDs, in principle half of the photons are lost due to the polarizers. Hence highly anisotropic structures used as a linearly polarized backlighting source can substantially reduce the light loss after the first polarizer, resulting in an enhanced and sharp image.

Two-dimensional (2D) structures can also exhibit polarized emission. A TEM image of 2D CdSe nanoplatelets is presented in Figure d. Nanoplatelets can be synthesized with an atomic precision over their thickness, resulting in an extremely narrow photoluminescence line width (Figure e), which can enhance the color purity in display applications. Additionally, nanoplatelets can also demonstrate highly directional emission.[12] Given the ability to overcome alignment obstacles for nanoplatelets arising from their 2D structure, their incorporation in LCD screens will also increase the observed brightness, due to minimization of photonic loss passing through the filters.

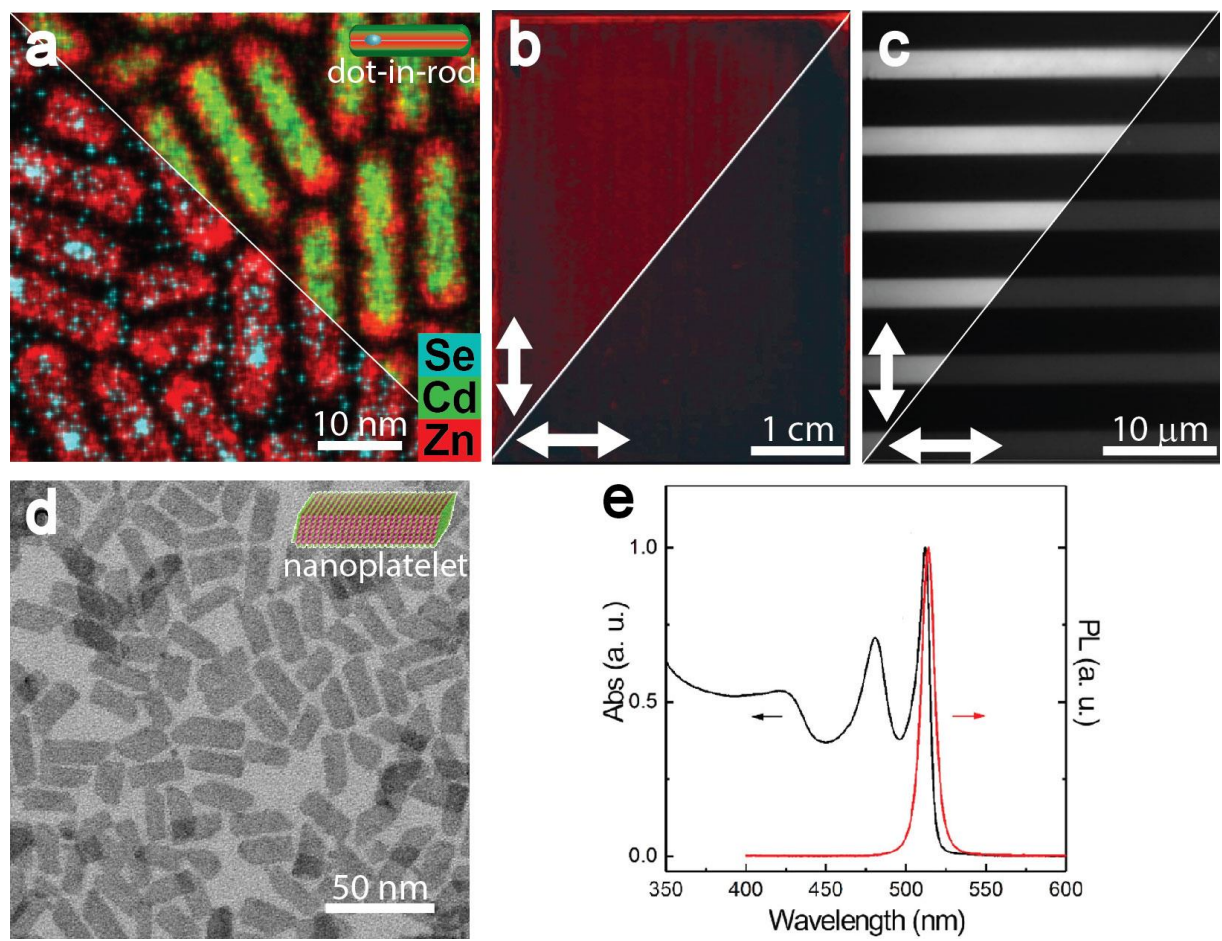

**Figure 3.** *The colloidal quantum materials palette.* a) Scanning transmission electron microscopy (STEM) image depicting elemental mapping of $CdSe/Cd_{1-x}Zn_xSe$ seeded nanorods with a graded shell; the upper triangle is an overlay of Zn and Cd spatial distributions, which demonstrates the graded-shell structure, while the bottom triangle is an overlay of Zn and Se spatial distributions, which demonstrates the seeded-rod structure. Adapted with permission from [8]. Copyright (2017) American Chemical Society. b) Polarized emission from a thin film of aligned CdSe/CdS nanorods, viewed through polarizers. Arrows indicate the orientation of the polarizer. c) Zoom in polarized microscope image of the interdigitated electrode set-up used to align the nanorods. Arrows indicate the orientation of the polarizer. d) TEM image of CdSe nanoplatelets with 6 monolayer thickness. e) Absorption (black) and photoluminescence (red) spectra measured on nanoplatelets with 4 monolayer thickness, exhibiting narrow photoluminescence linewidth. d) and e) adapted with permission from [13]. Copyright (2020) American Chemical Society.

## b. Chemical manipulation for incorporating colloidal quantum materials in displays

To incorporate SCNCs in displays, facile patterning methods must be applied, either for embedment within polymer films and also for patterning into the desired RGB pixels. We focus on a few important examples of patterning approaches and the interested reader is referred to a more extended review.[14]

A most applicable patterning method widely used in displays is photolithography, in which a UV light is shone on a substrate containing a polymer photoresist, through a mask containing the desired pattern. Upon illumination, the polymer changes its solubility, and the mask pattern is transferred to the substrate. Its most straightforward advantage is the ability to achieve high resolution patterns (<100 nm) onto a large area, relatively fast and inexpensively. Recently, direct optical lithography of functional inorganic nanomaterials (DOLFIN) was introduced, which is tailored specifically towards SCNCs.[15] In this approach, inorganic ligands which passivate the QDs, and dictate their solubility are replacing the role of the traditional photoresist. The ligand molecules, depicted in Figure a, are ion pairs termed as $Cat^+X^-$, in which $X^-$ is an electron-rich nucleophilic group, that binds to the Lewis acidic surface sites, usually the metal ions of the inorganic QD; the $Cat^+$ is the cation balancing the $X^-$. In non-polar media and on films, colloidal stability is afforded by the tight binding of the ion pair. Upon irradiation with UV light through a patterned mask, the photosensitive specie decomposes and depending on the choice of material, either the $Cat^+$ or the $X^-$, transforms the QDs insoluble in the developer solvent, which is usually a polar solvent used to wash-off the non-illuminated QDs. The procedure is illustrated in Figure b. Among the several possibilities for ion pairs to be used, they all must demonstrate dominant UV absorption bands, over the absorption spectra of the corresponding NCs. Figure c presents positive and negative patterns produced under the DOLFIN method. A variation in the QDs surface ligand allowed for the negative pattern tone.

Another facile and cost effective method is the transfer printing illustrated in Figure d. In this method, the desired pattern is printed on the substrate using an elastomeric stamp. The stamp, fabricated by lithographic means, is pressed upon a pad containing a thin layer of the QDs ink, which adhere to the protruding pattern on the stamp. The stamp is then brought into contact with the desired substrate, and pattern transfer occurs. The process is repeated with suitable QD inks, to complete the patterning of an RGB pixel. This method was used to successfully fabricate a 4-inch full-color active-matrix colloidal quantum dot (CQD) display with a resolution of 320×240 pixels by using CdSe/CdS/ZnS red-emitting CQDs, and CdSe/CdS green- and blue-emitting CQDs.[16] Transfer printing allows for large-area patterning, however, the method possesses a major drawback of deteriorating transfer quality with increasing resolution. The intaglio transfer printing method may resolve this problem.[17] As depicted in Figure e, a featureless elastomeric stamp is loaded with ink and pressed upon an engraved substrate. Upon detachment of the stamp, a mirror image of the engraved pattern is accurately left on the stamp, ready to be transferred onto the receiving substrate. The intaglio printing demonstrates superior performances, especially for high-resolution patterns as presented in Figure f for printing different pixel sizes. However, similar to the transfer method, it involves a cascade work plan which may introduce difficulties.

Ink-jet printing is an alternative patterning method applicable for the solution processable SCNCs. It is a simple, direct writing method, with a high degree of automation, which requires no mask and can produce complicated patterns at low costs. An electrically controlled nozzle head is used to drop a fixed volume of ink upon demand. Once the ink-drops hit the surface they laterally spread and dry to give a thin film. To improve the fairly low resolution of conventional ink-jet printing (10-20 $\mu$m), which is much below the photolithography or transfer created patterns, electrohydrodynamic (EHD) printing was introduced. A voltage bias is applied between a substrate and a metal-coated glass pipette, with a narrow nozzle, to draw a fine jet of ink through the nozzle (Figure g). Control over the thickness of the pattern is gained by a sequence of overlaid printing. Patterns with a resolution of ~50 nm are afforded and both, raster scan (Figure h) and drop on demand (Figure i) printing modes were demonstrated with good resolution.[18]

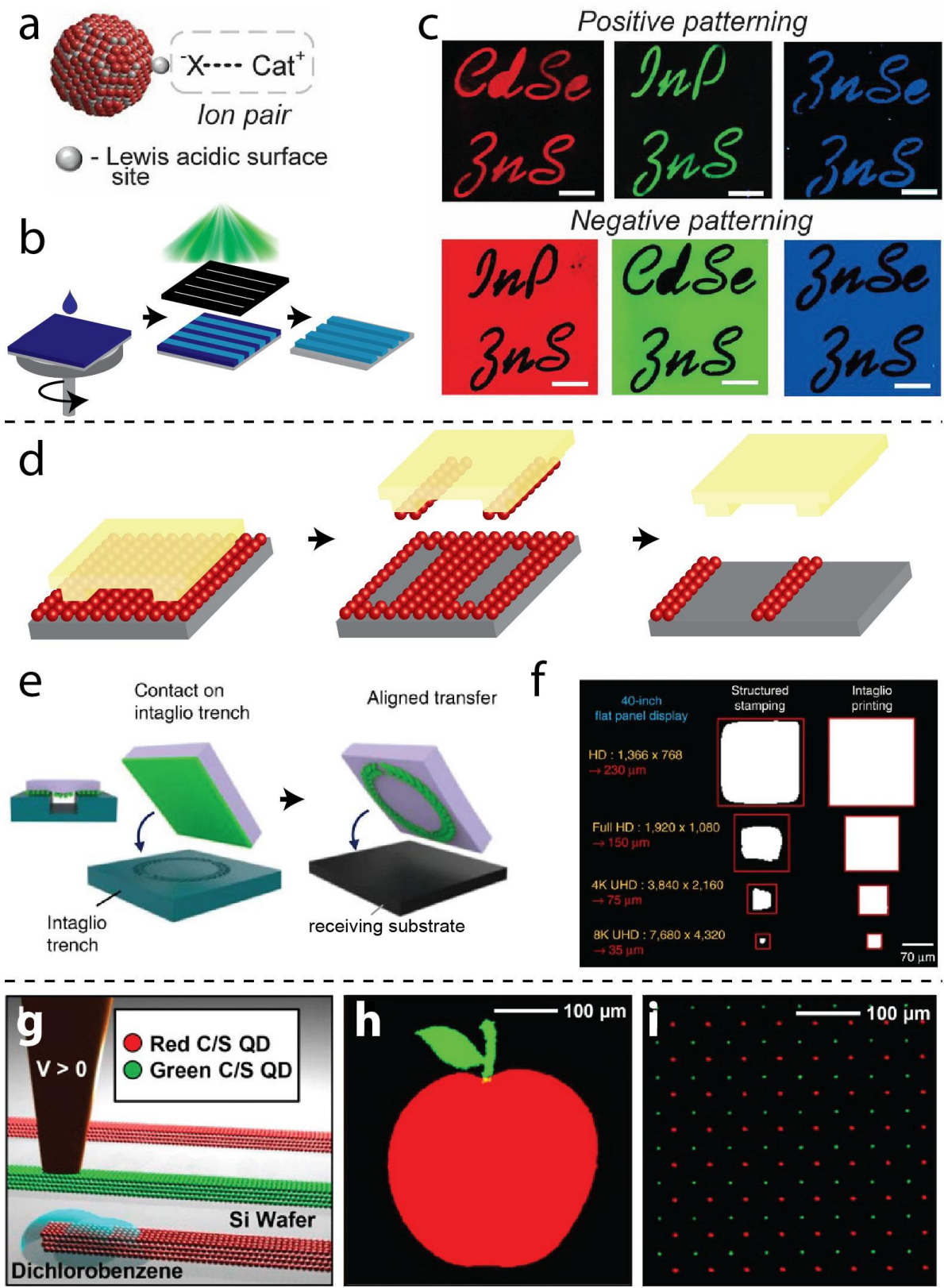

**Figure 4.** *Patterning methods.* a-c) DOLPHIN method, d-f) transfer printing methods, g-i) electrohydrodynamic ink-jet method. a) Scheme of the QD with the ion-pair surface ligand used in the DOLPHIN method, illustrated in b). c) Patterns of CdSe/ZnS (red), InP/ZnS (green), and ZnSe/ZnS (blue) core/shell QDs produced under the DOLPHIN method. a) and c) reprinted with permission from [15]. Copyright (2017) AAAS. d) Illustration of the transfer printing method. e) Schematic illustration of the intaglio transfer printing process. f) Comparison between the structured (left) and intaglio (right) stamping quality, with increasing pixel resolution. e) and f) adapted with permission from [17]. Copyright (2015) Springer Nature. g) Schematic illustration of a metal-coated glass nozzle and a target substrate during the printing process. h) Composite fluorescence images of a pattern of red and green QDs printed in a raster scanning mode to obtain uniform coverage. i) Composite fluorescence images of arrays of red and green QDs printed in a drop on demand mode. g)-i) reprinted with permission from [18]. Copyright (2015) American Chemical Society.

## 4. ROHS compatible Colloidal Quantum Materials

In 2006 the EU has restricted the use of cadmium in the Regulation on Hazardous Substances (or RoHS) directive, further intensifying the search for cadmium-free quantum dots. Yet, as innovative companies have been able to demonstrate the use of cadmium-based materials would result in significant $CO_2$ reduction and energy-saving, without a significant risk to health and environment, the EU released in 2009 an exemption for the use in displays, allowing further product development before its final ban in Europe from October 2019. This section elaborates on two of the main material families that are under development as alternatives, Indium-based and Zinc-based nanocrystals while mentioning additional material systems. Their spectral emission coverage is presented in **Error! Not a valid bookmark self-reference.**a.

InP quantum dots and their derivatives are the main materials that are developed to cover two out of the three corners in the trigonal color gamut in the CIE, the green, and red emissions. InP belongs to the III-V semiconductor family and its bulk bandgap is at 1.35 eV (918nm), allowing to tune its emission through quantum confinement effects over a wide range of the visible spectrum. Moreover, it has a strong blue absorption required with the blue backlight LED configuration. Yet, to date, it is typified by fairly wide emission peaks and the material is highly sensitive to oxidation, requiring the introduction of special synthetic solutions to address these limitations.

These two drawbacks may be addressed through a controlled synthesis of InP/ZnSe/ZnS core/shell system.[19,20] Stoichiometric control within both the core and shell regions achieved by purification of the precursors between the different shell growth stages, was found to result in InP/ZnSe/ZnS nanocrystals with a high quantum yield of 93% and QD-LED external quantum efficiency (EQE) of 12.2% (Figure 5b).[20] A follow-up study then demonstrated the achievement of highly spherical and symmetrical core/shell structure nanocrystals that further increased the stability and efficiency of the system.[21] The InP core size uniformity was increased by the use of a seeded growth approach, which separates the nucleation and growth steps. The oxidation was dealt with the use of hydrofluoric acid to etch any oxide surface. In parallel to the etching, a ZnSe interlayer was grown and its thickness was optimized to reduce Auger recombination and energy transfer which compete with the radiative recombination and reduce the quantum yield. Then, an outer ZnS shell was grown to passivate any surface traps and create a type I system with limited sensitivity to the environmental conditions. This resulted in narrow FWHM of 35nm, high quantum yield, theoretical QD-LED EQE of 21.4%, high brightness of 100,000 cd/m$^2$, and long operating half-life of 1,000,000 h at 100 cd/m$^2$.[21] Compared to the capacity of the InP-based QDs to provide the required features for the green and red pixels, a pure blue fluorescent InP-based material is still missing, and the current performances of the existing sky blue emitting InP QDs (>468nm) are still far behind.

Another family of materials which are investigated for the blue emission is Zinc chalcogenides and more specifically ZnSe and its derivatives. Zinc selenide belongs to the II-VI family of semiconductors and therefore resembles in some of its properties the cadmium-based systems. As zinc is smaller than cadmium, its bandgap is larger, resulting in a bulk bandgap of 2.7eV (460nm) making it a prominent candidate for the blue pixels. Moreover, doping it with Tellurium and a heterostructure conformation with ZnTe allow pushing its emission to cover also the green range if needed.[22] The relatively low quantum yield and high sensitivity towards oxidation of both ZnSe and ZnTe have restricted so far their potential use.

One way to solve these limitations is shell growth as for the cadmium and indium based quantum dots that would passivate surface traps and provide sufficient stability to reduce the need for oxygen and moisture barrier films within the displays. For example, the growth of a ZnS shell on ZnSe was shown to increase the quantum yield of the particles with the increase of the shell's thickness, from less than 1% up to 96%.[23] Interestingly, the procedure of shell growth and more specifically its rate, moving from thermodynamic to kinetic controlled growth was found to have a significant effect on the final structure and optical performances of the resulting QDs (Figure 5c-f). Under thermodynamic controlled growth, a more symmetrical structure was achieved, the maximal quantum yield was maintained in high shell thickness and the fluorescence blinking of single particles was significantly reduced compared to the kinetic growth mode, fast rate regime, in which the QY started to decrease from a specific thickness and irregular structures were formed. These differences were attributed to the avoidance of interfacial traps between the core/shell materials and uniform protection of the core from its sensitivity to the environment.

Additional systems, such as copper indium sulfide (CIS), metal halide perovskites, and bulk metal halides were also suggested and found to exhibit promising characteristics, such as narrow and tunable emission over a wide spectral range and potentially reduced manufacturing costs.[24] However, further work is required to optimize their performances for use in displays. For example, CIS exhibits low color purity (due to large FWHM) and relatively narrow color gamut, whereas at present the majority of halide perovskites and bulk metal halides still comprise heavy-metals and/or present low structural and chemical stability.

The introduction of the abovementioned novel synthetic routes has opened the path for the next generation of displays containing cadmium-free materials, fulfilling the immediate need resulting from the restrictions on the use of cadmium. However, the rarety and low yet existing toxicity of indium, the use of hazardous precursors, as well as the continuous development of other promising heavy-metal free materials, infer that newer and better nanocrystals for display applications are still desired and yet to come.

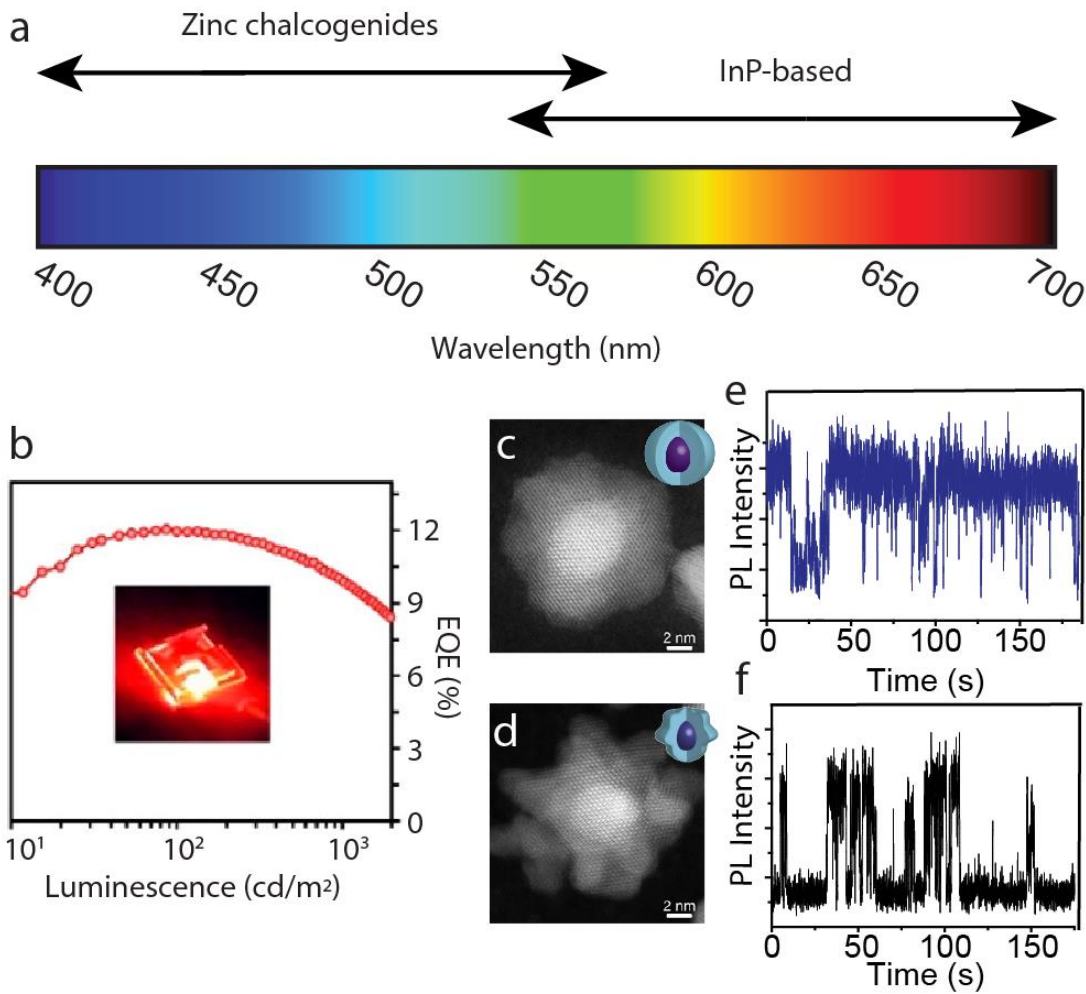

**Figure 5**. *Emerging cadmium-free QDs for display applications.* a) The spectral emission coverage of Indium-based and Zinc-based materials presenting the potential of the former for the green and red pixels and the latter for the blue and green. b) EQE-luminance profile for InP/ZnSe/ZnS QD-LED exhibits high EQE up to 12.2% which is maintained high also under 1000 cd/m$^2$. Inset, a photograph of a pixeled QD-LED. The figure was adapted with permission from Li et al.,[20] Copyright (2019) American Chemical Society. c-f) The effect of ZnS shell growth rate on the morphology, and blinking of ZnSe/ZnS core/shell nanoparticles. c-d) TEM images of the ZnSe/ZnS QDs synthesized by thermodynamic shell growth mode (slow rate) reveal a symmetrical structure (c) compared to irregular island growth for the kinetic growth (fast rate, d). e-f) The fluorescence blinking profile for single particles of the two structures reveals that the thermodynamic growth mode, leads to much less blinking (e) compared to the kinetic growth mode (f). Figures 5c-f were adapted with permission Ji. et al.,[23] Copyright (2020) American Chemical Society.

## 5. Outlook for colloidal quantum materials in displays

In the previous section, the material challenges for SCNCs in displays were presented, especially the search for Cd-free materials in light of the ROHS regulations. Moreover, current SCNCs offer new possibilities for future displays. In the following section, we will outlook the possible new modes of operations which can enhance the efficiency and brightness of QD and simplify the architecture of the displays, paving the way for thinner and flexible displays. Then, we will outlook new applications which can revolutionize our everyday life, in which transparent, flexible, and multi-functional SCNCs displays may contribute to.

**i. New modes of operation**

In the first section, the simple "Drop-in" solution of QDs enhancement films in LCDs was discussed, which does not require the LCD manufacturers to change their highly advanced and complex production processes. However, as mentioned, the challenges of this operation mode are reduced efficiency and brightness. These issues mostly arise from the fact that the red, green, and blue lights are generated in the back light unit which are then filtered in the 3 sub-pixels color filters. Another source of losses arises from the use of spherical SCNCs. As written above, these SCNCs emit unpolarized light which is inevitably filtered in the liquid crystal unit polarizers. Besides, the emission is isotropic with no preferred direction. As a consequence, only a small part of the emitted photons are doing their way out of the display, and many efforts are given to address these issues. In fact, this light outcoupling challenge is a limitation also for OLED displays in which the record EQE is less than 40%[25].

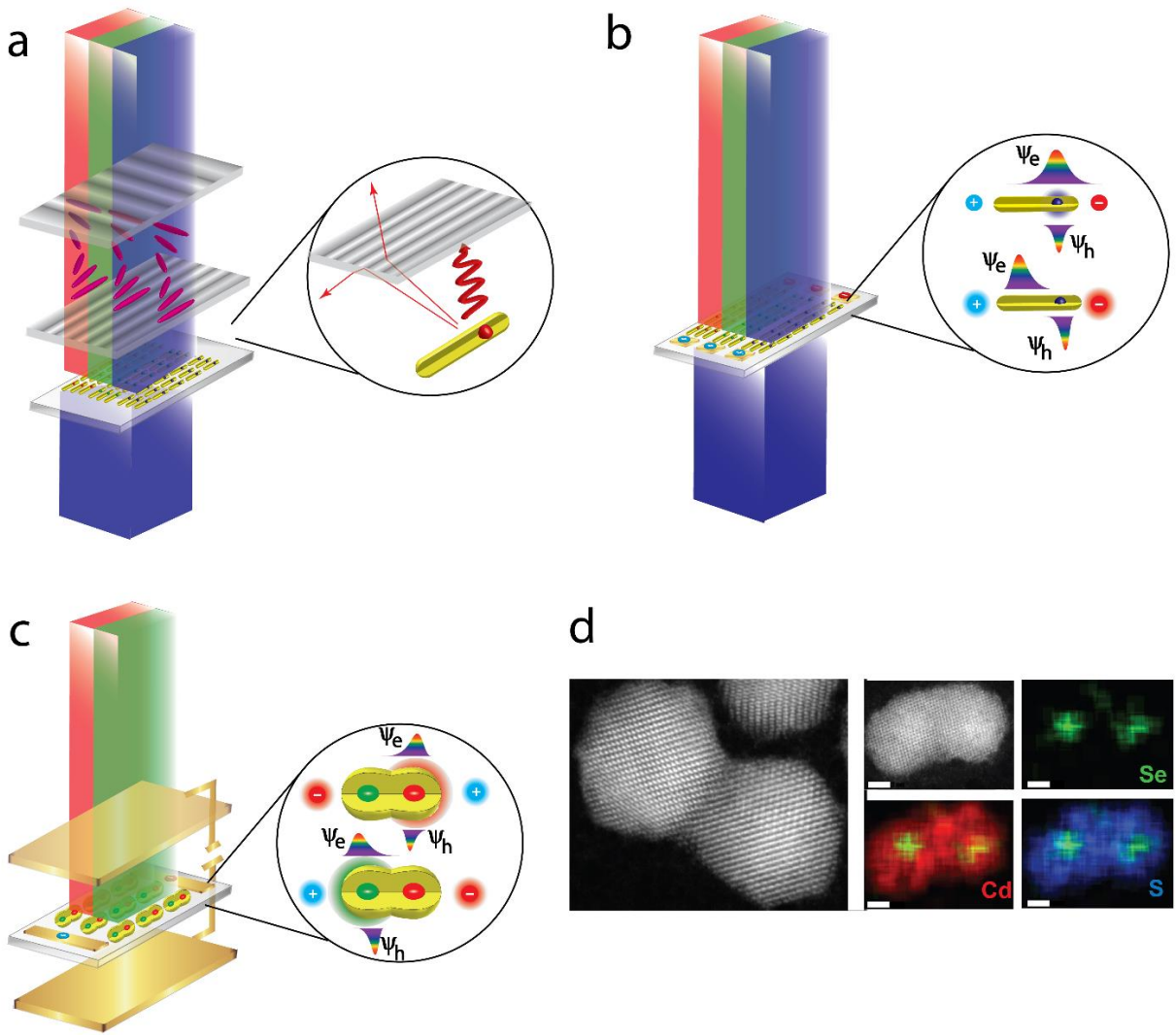

**Figure 6.** *New modes of operation for SCNC QD displays.* a) Incorporation of aligned dot-in-rods inside an enhancement film. b) Light modulation by electric field along the long axis of dot-in-rods. c) Electric field dependent color of a pixel by CQDMs, d) HRTEM, and EDS elemental analysis images of CQDMs. reprinted with permission from[26]. Copyright AIP Publishing.

As discussed above, CdSe/CdS dot-in-rod are known to be highly efficient emitters of polarized light which can reach emission polarization values around 0.8. Their incorporation, while aligned, as an enhancement film in the BLU may make the simple "Drop-in" solution more efficient. Their blue absorption is far less polarized than their emission, allowing them to absorb light which cannot make its way through the vertical polarizer by emitting this light in the correct polarization (Fig. 6a). In addition, since the emission is polarized, we can treat the dot-in-rod as a dipole emitter in which the angular distribution of emission is more concentrated orthogonal to the dipole axis allowing for more light to impinge on the outer surface of the display at an angle less than the critical angle for total internal reflection, allowing more photons to be directed out of the display (Fig. 6a).

Thinking of thin and flexible displays, electroluminescent QD displays could be made thin and flexible but, as discussed above, still suffer from low QY and stability issues compared to photoluminescent displays. Therefore photoluminescent displays, in which the modulation of every sub-pixel color is made by an electric field, might be an alternate solution. Early attempts on on/off switching of CdSe nanorods, aligned with their long axis with the electric field direction, showed potential for this direction[27]. This was followed by experiments on CdSe/CdS dot-in-rods with quasi-type-II band alignment, in which the hole wave-function is localized in the CdSe core, while the electron wave-function is delocalized over all the rod[28]. Under applied electric field parallel to the long axis of the nanorods, the electron and hole wave-functions are being separated leading to decreased emission (Fig. 6b). Since every sub-pixel color can be modulated separately by the electric field, any color composed out of the red, green, and blue base colors can be generated.

Furthermore, what if the color of the subpixel can be determined by the electric field? New emerging CQDM (Coupled QD Molecules) suggest an innovative approach in this direction[26,29]. These CQDMs are made by fusion of two CdSe/CdS core/shell nanocrystals building-blocks resulting in a continuous nanocrystal molecule (Fig. 6d). Elemental analysis shows that after fusion the two CdS shells are becoming a rod-like nanocrystal with two emission centers at the two CdSe cores (Fig. 5d). CQDMs with two different cores have the ability of dual-color emission, red and green for example. Aligning them with their long axis parallel to the electric field direction has the potential to dictate the emission color (Fig. 6c). This layer can be inserted as the emissive layer in an electroluminescent display architecture.

## ii. Future applications for QD displays

The ever-lasting shrinking thickness of QD displays especially electroluminescent displays opens up vast new applications where QD displays take a major role. Here we mention only a few applications in which some progress was already reported, and we envision it will lead to a significant change in our everyday experiences.

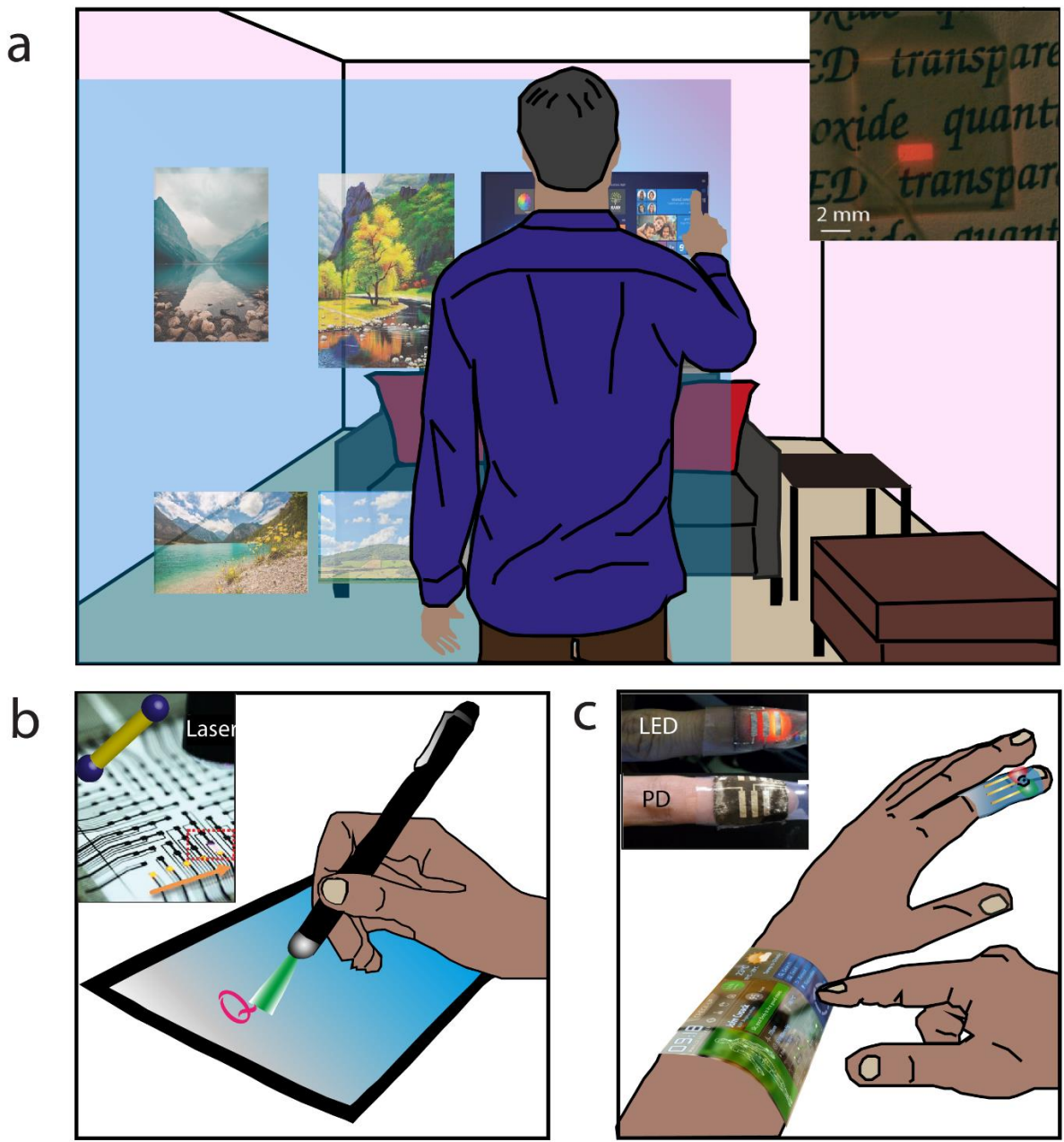

**Figure 7. *Future applications for QDs displays.*** a) Transparent QD displays can be integrated into glasses or windows allowing projection of information on the background. Inset reprinted with permission from[30]. Copyright (2010) American Chemical Society. b) Multi-functional DHNRs enabling tablets with stylus pen made out of laser pointer. Inset reprinted with permission from[31]. Copyright (2017) AAAS. c) Multi-functional displays. QD displays are integrated with other electronic components like touch sensors or bio-sensors enabling wearable flexible electronic bracelets. Inset reprinted with permission from[32]. Copyright (2017) American Chemical Society

Transparent QD displays can be integrated into glasses or windows allowing projection of information on the background without affecting it. In the future, such windows can replace the conventional artworks hanged on walls, in which every image captured with our smartphone, can be immediately presented on the windows. (Fig. 7a).

So far the performance of transparent QD displays is inferior to their non-transparent counterparts. This stems from the fact that the transparent layer puts another constraint on the materials available with suitable energy levels for electroluminescent displays. However, transparent QD displays were already demonstrated with ITO electrodes in field-induced QD ionization architecture[30] (inset Fig. 7a).

Another promising application for QD displays can emerge from multifunctional SCNCs. This was successfully demonstrated using DHNRs (Double-heterojunction nanorods) in a dumbbell structure made out of CdS rods with CdSe tips covered with ZnSe. These nanocrystals, as an emissive layer in the electroluminescent device, were shown not only to emit light under direct bias but also to generate photocurrent under reversed bias[31]. These particles can be integrated into tablets or note smartphones where the stylus pen could be a laser pointer that impinging on the pixels in reverse bias, which are then emitting under direct bias (Fig. 7b and inset).

Another area where flexible QD displays are already making their first steps is in multifunctional displays. In these systems, the display is integrated with other electronic components like touch sensors arrays for user input which is then presented on the display. These can emerge to be a flexible wearable electronic devices which can be wrapped around the hand[33] (Fig. 7c). In this context, flexible transparent displays can be also integrated with biosensors made themselves from flexible transparent QD LEDs. As an example a stretchable QLED and photo-detector), made also from QDs, can be wrapped around the fingertip to measure the photoplethysmogram (PPG) signal and then wirelessly transmit the signal to the QD bracelet[32] (Fig. 7c and inset).

This review explored the current status of QD displays and their roadmap to future QD displays. The vast advantages and diversity of colloidal semiconductor nanocrystal QDs for displays were presented, along with the major challenges to future displays, especially the search for Cd-free materials. Alternative materials were also suggested, some are already doing their way into commercial displays. We hope that our suggested outlook for new architectures and applications, in which SCNCs might be incorporated as displays, will stimulate scientists and entrepreneurs for developing even better and more innovative utilization of SCNCs in displays.

**Biographies:**

*Yossef E. Panfil* completed his BSc at JCT and then his MSc in Applied Physics at the Hebrew University of Jerusalem. He is currently a PhD student at the Hebrew University of Jerusalem in Prof. Banin's group. His research focuses on coupling effects in coupled colloidal Semiconductor QD molecules, using a combination of novel single nanocrystal spectroscopy methods and numerical computations.

*Meirav Oded* received her Ph.D. (2016) in Chemistry from the Hebrew University of Jerusalem under the guidance of Prof. Roy Shenhar. During her Ph.D. she studied selective deposition of polyelectrolytes over block copolymer templates with nanometric resolution. She is currently a staff scientist and lab supervisor at the Prof. Banin group. Her research interests are surface chemistry and self-assembly methods of colloidal semiconductor nanocrystals.

*Dr. Nir Waiskopf* is a team leader, who manages the photocatalytic research in Banin's group. He holds a BSc degree in physics and biology from the Hebrew University of Jerusalem as well as a MSc (magna-cum-laude) and PhD in chemistry and nanotechnology, on nano-based systems for diverse biomedical applications.

*Professor Uri Banin* holds the Alfred & Erica Larisch Memorial Chair at the Institute of Chemistry and the Center for Nanoscience and Nanotechnology at the Hebrew University of Jerusalem (HU). Banin obtained his PhD in Physical Chemistry (Magna Cum Laude, 1995) from the Hebrew University of Jerusalem and performed his postdoctoral research at the University of California at Berkeley (1994-97). He was the founding director of the HU Center for Nanoscience and Nanotechnology (2001–2010) and is a founder of Qlight Nanotech that developed colloidal quantum materials for display applications, and was acquired by Merck KGaA, Darmstadt, Germany. His research focuses on the chemistry and physics of nanocrystals including synthesis of nanoparticles, size- and shape-dependent properties, and emerging applications in displays, lighting, solar energy harvesting, 3D printing, electronics, and biology.

**Acknowledgments**

Funded in part by the European Research Council (ERC) under the European Union's Horizon 2020 research and innovation programme (grant agreement No [741767]). U.B. thanks the Alfred & Erica Larisch memorial chair. Y.E.P. acknowledges support by the Ministry of Science and Technology & the National Foundation for Applied and Engineering Sciences and the Council for Higher Education, Israel.